\RequirePackage{fix-cm}
\documentclass[smallextended]{svjour3}       
\smartqed  
\usepackage{graphicx}
\usepackage{amsmath,bm}
\usepackage{amsfonts}
\begin{document}

\title{Bayesian Filtering for Multi-period Mean-Variance Portfolio Selection
}

\author{Shubhangi Sikaria \and
        Rituparna Sen \and
		Neelesh S. Upadhye         
}

\institute{S. Sikaria \at
              Indian Institute of Technology, Madras, India \\
              Tel.: +9176816664\\
              \email{shubhangisikariya@gmail.com}           
           \and
           R. Sen \at
           Indian Statistical Institute, Bangalore, India
}

\date{Received: date / Accepted: date}

\maketitle

\begin{abstract}
For a long investment time horizon, it is preferable to rebalance the portfolio weights at intermediate times. This necessitates a multi-period market model. Usually, dynamic programming techniques are applied to optimize the portfolio for the multi-period model. However, this assumes a known distribution for the parameters of the financial time series. We consider the situation where the distribution of parameters is unknown and is estimated directly from the dynamically arriving data. We implement the Bayesian filtering method through dynamic linear models to sequentially update the parameters. We also acknowledge the uncertain investment lifetime to make the model more adaptive to the market conditions. These updated parameters are put into the dynamic mean-variance problem to arrive at optimal efficient portfolios. Implementing this model to the S\&P500 illustrates that the data strongly favor the Bayesian updating and is practically implementable.

\end{abstract}

\section{Introduction}
\label{sec:1}
Investing in the stock market exposes the investors to the risk, which can be reduced by investment in an assortment or range of securities. The major challenge faced by investors is how to allocate their capital over many financial assets. Thus, the intention is to ascertain an optimal portfolio that gives the best allocation of wealth by generating a high return along with a low risk. Markowitz \cite{markowitz1952portfolio} paved the foundation of the modern portfolio theory, which is modeled as a return-risk bi-criteria optimization problem, characterizing portfolio return with a mean rate of return and risk with variance. The analytical solution of mean-variance portfolio selection (MVPS) in a single-period was derived by Merton \cite{merton1972analytic}. However, the single-period model is static, which implies that the decision can be made only at the beginning of the investment period, and one needs to wait for results until the investment horizon ends. Due to the long investment horizon, it is preferable to rebalance the portfolio weights at intermediate time points after observing current values. A natural extension to a multi-period model allows investors to make financing decisions at distinct time points, and account for the immediate market scenario. Among multi-period investment models, one can have the time horizon to be continuous and discrete. Zhou and Li \cite{zhou2000continuous} formulated a closed-form solution in a multi-period model for a continuous-time setting by embedding nonstandard problems into a class of auxiliary stochastic linear-quadratic (LQ) problems. On a similar line, Li and Ng \cite{li2000optimal} proposed an efficient algorithm for determining an optimal analytical solution to maximize utility function in the multi-period discrete-time portfolio selection problem.

In work mentioned above, the stock prices are assumed to follow a vector-valued Geometric Brownian motion implying that the stock returns are time-independent. The empirical evidence shows that the returns of risky assets always exhibit serial correlations that are captured by ARMA models. Balvers and Mitchell \cite{balvers1997autocorrelated} were first to derive an explicit analytical solution to the dynamic portfolio problem by employing normal ARMA(1,1) to incorporate the autocorrelation among returns of risky assets. Xu and Li \cite{xu2008dynamic} investigated dynamic portfolio selection for serially correlated returns by embedding the mean-variance model into a quadratic utility model. Later, they applied dynamic programming for one risky asset and one risk-free asset to obtain the explicit optimal investment strategy. General forms of correlation structure for returns based on stochastic market for single risky asset was assumed by  \c{C}elikyurt and \"{O}zekici \cite{celikyurt2007multiperiod} and Dokuchaev \cite{dokuchaev2007discrete}. Gao et al. \cite{gao2015time} and Chiu and Wong \cite{chiu2014mean} investigated a dynamic MVPS for multiple risky assets and one riskless asset with a general correlation for a market. He and Wang \cite{he2015multi} established the optimal investment policy for the multi-period model by solving the stationary equation directly without using the embedding technique. Meanwhile, to make the MVPS model more realistic the multi-period discrete-time model has been studied extensively by incorporating various real features in recent years (see the ref. \cite{gao2014multiperiod}, \cite{wei2015dynamic}, \cite{skaf2009multi}).

Originally, the parameters like drifts and volatilities of returns for the portfolio selection models are estimated from the past data and remain constant for later periods. However, this solution is not realistic as the parameters are not adaptive according to market conditions. To circumvent this issue, Mao and Sarndal \cite{mao1966decision} used Bayesian inference for a discrete-time single period portfolio selection model. Later an extensive literature incorporating Bayesian statistics have emerged, see Frost and Savarino \cite{frost1986empirical} and Aguilar and West \cite{aguilar2000bayesian}. Karatzas and Zhao \cite{karatzas2001bayesian} did the most recognizable work on Bayesian learning to compute optimal portfolio allocation for an unknown drift and Gaussian asset returns. Recently, Gu'eant and Pu \cite{bismuth2019portfolio} extended the previous results for optimal portfolio liquidation and transition problems in continuous time in which expected returns of risky assets are estimated online. Franco et al. \cite{de2018bayesian} adapted the methodology of Zhou and Li \cite{zhou2000continuous} to the Bayesian learning framework and embedded the Bayesian-Markowitz problem into an auxiliary standard control problem and then applied dynamic programming approach. Bodnar et al. \cite{bodnar2017bayesian} deal with the global minimum variance portfolio problem where the prior distribution of logarithmic returns are assumed to be normal and independent. They utilized various standard priors for mean vector and covariance matrix to derive posterior distribution for the weights. Recently, Bauder et al. \cite{bauder2018bayesian} derived posterior prediction distributions of returns to obtain optimal portfolio weights by assuming returns to be infinitely exchangeable and multivariate centered spherically symmetric for unknown mean vector and covariance matrix. In the literature mentioned above, asset returns are independent of past observations. However, empirical evidence has shown a serial correlation among the financial time series, which we captured in this paper using the vector autoregressive (VAR) model. Later, we applied linear Bayesian filtering approach to update the VAR model parameter sequentially and to obtain posterior prediction distribution.

The literature mentioned above makes an implication that the investment time horizon is deterministic, that is investor operates the investment strategy until the explicit exit time. However, due to market situations or an investor's personal reasons, he/she may be forced to leave the financial market before the exit time. In that case, the investment lifetime is uncertain and considered as a random variable. Martellini and Uro\u{s}evi\'{c} \cite{martellini2006static} maximize the quadratic expected utility function with uncertain exit time in which exit time depends on asset price behavior. Guo and Hu \cite{guo2005multi} analyzed a multi-period mean-variance investment problem with an uncertain time of exiting. Later, some researchers \cite{huang2008portfolio}, \cite{blanchet2008optimal}, \cite{yi2008multi}, \cite{huang2010portfolio} extended the portfolio selection problem with stochastic time horizon by adopting efficient methodologies to make it more practical/realistic. Zhang and Li \cite{zhang2012multi} derived an analytical solution for a multi-period optimization problem with serially correlated returns and considered uncertain exit time as an exogenous random variable with a discrete probability distribution.

We consider the mean-variance portfolio optimization problem with uncertain exit time and serially correlated returns of multiple risky assets whose parameters are updated dynamically using Bayesian forecasting for dynamic linear models. We assume that the distribution of exit time is discrete and known, as it was in Zhang and Li \cite{zhang2012multi}, and the returns of the risky asset are autocorrelated subject to normal VAR(p) process. Firstly,  we convert the autoregressive model into a dynamic linear model (DLM) whose parameters are assumed to follow a normal distribution with mean and variance estimated from the past data. DLM has been used previously in financial optimization problems by Irie and West \cite{irie2019bayesian}, where the returns are still considered to be independent over time. To sequentially update the distribution and to obtain a one-step-ahead forecast of returns and posterior distribution of VAR model parameters, we apply Bayesian techniques by West and Harrison \cite{west2006bayesian}. We implement a dynamic programming approach for an uncertain time horizon to the updated returns to obtain the optimal strategy for our investment.

The rest of the paper is organized as follows. Section \ref{sec:3} illustrates the dynamic linear model and finds the prior and posterior distributions for the AR parameters.  Section \ref{sec:2} summarizes a dynamic programming approach to solve the mean-variance multi-period portfolio selection problem with uncertain exit time. Section \ref{sec:4} generalizes the model to multiple assets. Section \ref{sec:5} investigates the dependence of the return series on the specific AR parameter values. Section \ref{sec:6} presents the empirical results of comparing the efficient frontiers obtained from the dynamic mean-variance portfolio selection problem with sequentially updating the parameters and fixed parameters. Section \ref{sec:7} presents the concluding remarks.

\section{Bayesian Forecasting}
\label{sec:3}
Prior research has found that the estimation of parameters using a Bayesian framework can improve the constructed portfolio's performance. In this work, we also take the Bayesian approach to account for estimating the risk of a portfolio in predicted stock returns. For details and proofs, refer to West and Harrison \cite{west2006bayesian}. Here, we consider that the returns are serially correlated and fit autoregressive models to estimate the underlying time series of returns. The returns of the risky asset at time period $t (t=1,..., T)$ within the investment horizon is denoted by $r_t$. Suppose that $r_t$ is a weakly stationary AR($p$) series given by
\begin{equation}\label{eq:(5)}
r_t = \mu + \sum_{i=1}^p \phi_i(r_{t-i}-\mu) + \epsilon_t,
\end{equation}
for some sequence of coefficients $\phi_1,\phi_2,...,\phi_p$ where $\mathrm{E}[r_t|\mathcal{F}_t] = \mu$ and the $\epsilon_t$ are zero-mean, uncorrelated random quantities with constant variance $\sigma^2$.

Consider a dynamic linear model with $r_t$ a $T$-vector based on $T$-vector $\theta_t$ via:
\begin{equation}
\mathbf{r}_t= \mathbf{F}_t'\mathbf{\theta}_t + \mathbf{\nu}_t, \hspace*{9mm}\mathbf{\nu}_t\sim N[0, \mathbf{V}_t]
\end{equation}
\begin{equation}
\mathbf{\theta}_t= \mathbf{G}_t\mathbf{\theta}_{t-1} + \mathbf{\omega}_t, \hspace*{5mm}\mathbf{\omega}_t\sim N[0, \mathbf{W}_t]
\end{equation}
\begin{equation*}
(\mathbf{\theta}_0|D_0) \sim N[\mathbf{m}_0, \mathbf{C}_0]
\end{equation*}
for some prior moments $\mathbf{m}_0$ and $\mathbf{C}_0$ and where the $\mathbf{\nu}_t, \mathbf{\omega}_t$ are independent and mutually independent innovations sequences.

All autoregressive models can be written in dynamic linear form in a variety of ways for various purposes. Here, we are considering regression model form to write AR($p$) model of equation \eqref{eq:(5)} as a simple, static regression sequentially defined over time
$$r_t = \mathbf{F}_t' \theta + \nu_t,$$
where $\mathbf{F}_t' = (r_{t-1},...,r_{t-p})$, $\theta' = (\mu,\phi_1,...,\phi_p)$ and $\nu_t = \epsilon_t$. Now the posterior distributions for AR parameters can be obtained by employing standard DLM results.The standard normal theory assumes that initial values $\mathbf{F}_1' = (r_{0},...,r_{-p+1})$ is known at the origin $t=0$. Also initial information to implement DLM form is as follows:
$$(\mathbf{\theta}|D_0) \sim N[\mathbf{m}_0, \mathbf{C}_0].$$
where $\mathbf{m}_0 = \mathrm{E}[\theta|D_0]$ and $\mathbf{C}_0 = \mathrm{Var}[\theta|D_0]$.

Sequential learning of AR parameters can be obtained by employing standard DLM results. As the AR parameters $\theta$ is constant for a time period, posterior distribution of $\theta$ at time $t-1$ given by $(\theta|D_{t-1})\sim N(m_{t-1},C_{t-1})$ concides with prior distribution at time $t$. Thus, prior distribution of $\theta$ at time $t$ is $(\theta|D_{t})\sim N(m_{t-1},C_{t-1})$.

One step ahead prediction distribution of the return series can be obtained from the prior distribution of $\theta$ which would lead to 
\begin{align}\label{eq3}
&(r_t|D_{t-1})\sim N(f_t, Q_t)\\
\nonumber\text{where, }& f_t =\mathbf{F}_t' \mathbf{m}_{t-1},\\
\nonumber & Q_t = \mathbf{F}_t'\mathbf{m}_{t-1}\mathbf{F}_t + \sigma^2.
\end{align}

Posterior distribution of $\theta$ at time $t$ can be derived from the prediction distribution of the returns, which is:
\begin{align}\label{eq4}
&(\theta_{t-1}|D_{t-1})\sim N(m_{t},C_{t})\\
\nonumber \text{where, }& \mathbf{m}_{t} = \mathbf{m}_{t-1}+ \mathbf{A}_t (r_t - f_t),\\
\nonumber &\mathbf{C}_t = \mathbf{C}_{t-1}-\mathbf{A}_t Q_t \mathbf{A}_t', \hspace{5mm} \mathbf{A}_t= \mathbf{C}_{t-1}\mathbf{F}_t Q_t^{-1}.
\end{align}
As $\theta_t$ is constant for a time period so this posterior distribution can be used as prior to forecast the returns.

\section{Dynamic Programming}
\label{sec:2}
In a multi-period setting, an investor can reallocate the portfolio's weights at intermediate times after observing the values until that time. Due to unpredictability in the financial market, it is always more profitable to evolve the portfolio after observing the current market scenario. As defined by Iyengar \cite{iyengar2005robust}, Dynamic programming (DP) is the mathematical framework that permits the decision-maker to compute a decent overall strategy efficiently by evolving data state. In discrete-time portfolio optimization, the investor can rebalance his wealth at distinct points in time with the most recent information.

Consider that a capital market consisted of a risky security with random returns and a riskless security with deterministic returns. An investor with an initial wealth of $x_0$ joins the market at time $0$ and allocates his/her wealth among these two securities. The investor can reallocate his/her wealth invested in the risky asset at the beginning of each consecutive time period. The updated returns of the risky asset obtained after employing Bayesian forecasting at time period $t (t=0,1,2,...,T)$ are denoted by a vector $\Tilde{r}_t$, which are considered as serially correlated, that is, the value of $\Tilde{r}_t$ depends upon the past realized returns $\Tilde{r}_s, s<t$. Let $r_t^0$ be the given deterministic return of the riskless security at period $t$. The uncertain exit time $\tau$ is considered as a discrete random variable and the actual exit time is $T\wedge\tau = min\{T,\tau\}$ with probability distribution $p_t = \mathbb{P}\{T\wedge\tau =t\}, t=1,2,...$.

Define the excess return of a risky asset at period $t$ as $e_t=\Tilde{r}_t-r_t^0$. We use the updated return series to calculate the mean and mean square for excess returns for $t= 1,2,...,T$
\begin{align*}
\mathrm{E}[e_t] &= \mathrm{E}[\Tilde{r}_t -r_t^0] = \mathrm{E}[\Tilde{r}_t]-r_t^0 = f_t -r_t^0,\\
\mathrm{E}[e_t^2] &= \mathrm{E}[(\Tilde{r}_t-r_t^0)^2]\\
& = \mathrm{E}[\Tilde{r}_t^2]+ (r_t^0)^2 -2r_t^0 \mathrm{E}[\Tilde{r}_t]\\
& = Q_t + f_t^2+ (r_t^0)^2 -2r_t^0 f_t.
\end{align*}
where $f_t = \mathrm{E}[e_t]$ is mean of excess return and $Q_t = \mathrm{E}[e_t^2]$ is mean square of excess return. Let $x_t$ be the wealth of the investor at the beginning of each time period $t (t=0, 1,..., T)$. Define investment series over T periods, $U:=\{u_0, u_1,..., u_{T-1}\}$, where $u_t$ is the amount invested in the risky security at time $t$. The investment strategy $U$ is assumed to be self-financing, that is, there is no exogenous inclusion or exclusion of money, which can be described mathematically as:
\begin{equation}\label{eq1}
x_{t+1}=\Tilde{r}_t^0 x_t + e_t u_t, \hspace{5mm}t=0,1,...,T-1.
\end{equation}
The multi-period MVPS for uncertain exit time can be written as follows
\begin{equation}
P(\omega)\left\{ \begin{array}{ll} \underset{U}{\text{max}} \hspace{3mm} \mathrm{E}_0(x_{T\wedge\tau})-\omega \mathrm{Var}_0(x_{T\wedge\tau}) \\ \text{s.t.} \hspace{5mm} x_{t+1}=r_t^0 x_t + e_t u_t, \hspace{5mm}t=0,1,...,T-1,\end{array}\right.
\end{equation}
where $\omega$ is a given positive constant, representing the investor's risk aversion and illustrates the attitude of investors towards risk.

Due to nonlinearity of conditional variance, employing dynamic programming to this multi-period portfolio optimization is not straight forward. More precisely, for the expected value operator, dynamic programming is applicable because of the smoothing property: $\mathrm{E}[\mathrm{E}(\cdot|\mathcal{F}_j)|\mathcal{F}_k] = \mathrm{E}(\cdot|\mathcal{F}_k)\: \forall j>k$, where   $\mathcal{F}_t$  denotes an information set available at time $t$ and $\mathcal{F}_{t-1}\subset \mathcal{F}_t, \forall t$, while the variance operator does not satisfy the smoothing property: $\mathrm{Var}[\mathrm{Var}(\cdot|\mathcal{F}_j)|\mathcal{F}_k] \neq \mathrm{Var}(\cdot|\mathcal{F}_k) \:\forall j>k$. Thus, Li and Ng \cite{li2000optimal}  transformed the original mean-variance equations into an auxiliary framework to find an analytical optimal solution which is described in the appendix. Here we present the optimal portfolio allocation strategy and efficient frontier for the original problem $P(\omega)$. 

\begin{theorem}
The optimal strategy of the mean-variance problem $P(\omega)$ is given by
\begin{equation}
u_t^* = \frac{1+2\omega\lambda_0 x_0}{2\omega(1-\Theta)}\frac{\mathrm{E}_t(\lambda_{t+1}e_t)}{\mathrm{E}_t(\omega_{t+1}e_t^2)} - \frac{\mathrm{E}_t(\omega_{t+1}e_t)}{\mathrm{E}_t(\omega_{t+1}e_t^2)} r_t^0 x_t,
\end{equation}
where
\begin{align*}
\theta_t &= \frac{\mathrm{E}_t^2(\lambda_{t+1}e_t)}{\mathrm{E}_t(\omega_{t+1}e_t^2)}, \hspace{15mm} \Theta = \sum_{t=1}^T \mathrm{E}_0(\theta_{t-1}),\\
\omega_t &= p_t+(r_t^0)^2\left[\mathrm{E}_t(\omega_{t+1}) - \frac{\mathrm{E}_t^2(\omega_{t+1}e_t)}{\mathrm{E}_t(\omega_{t+1}e_t^2)}\right], \hspace{5mm} \omega_T = p_T,\\
\lambda_t &= p_t+r_t^0\left[\mathrm{E}_t(\lambda_{t+1}) - \frac{\mathrm{E}_t(\omega_{t+1}e_t)\mathrm{E}_t(\lambda_{t+1}e_t)}{\mathrm{E}_t(\omega_{t+1}e_t^2)}\right], \hspace{5mm} \lambda_T = p_T,\\
\end{align*}
for $t=0,1,...,T-1.$

The efficient frontier of the original problem $P(\omega)$ is given by
\begin{equation}
\mathrm{Var}_0(x_{T\wedge\tau})=\frac{(1-\Theta)}{\Theta}\left[\mathrm{E}_0(x_{T\wedge\tau})-\frac{\lambda_0 x_0}{1-\Theta}\right]^2 + \left[\omega_0-\frac{\lambda_0^2}{1-\Theta}\right]^2x_0^2, \label{eq:(2)}
\end{equation}
for $\hspace{5mm} \mathrm{E}_0(x_{T\wedge\tau}) \in \left[\frac{\lambda_0 x_0}{1-\Theta},+\infty\right).$
\end{theorem}
Note that $\omega_{t+1}$ and $\lambda_{t+1}$is known at time $t$, for $t=0,1,...,T-1$, so
$$\mathrm{E}_t^2(\omega_{t+1}e_t)=\omega_{t+1}\mathrm{E}_t^2(e_t)\text{      and      }\mathrm{E}_t(\omega_{t+1}e_t^2)=\omega_{t+1}\mathrm{E}_t(e_t^2),$$
$$\mathrm{E}_t^2(\lambda_{t+1}e_t)=\lambda_{t+1}\mathrm{E}_t^2(e_t)\text{      and      }\mathrm{E}_t(\lambda_{t+1}e_t^2)=\lambda_{t+1}\mathrm{E}_t(e_t^2).$$

\section{Generalized Model for Multiple Risky Assets}
\label{sec:4}
We extend the previous portfolio optimization problem to the general situation with multiple risky assets. We model the returns of the risky assets using a vector autoregressive model of order $p$ (VAR$(p)$) to account for the serial correlation within the returns of a risky asset and the cross-correlation among the returns of different risky assets. Here, we consider $n$ securities with random returns $r_t^i \:(i = 1,2, \cdots,n)$  and $t=(1,2,\cdots,T)$ and a riskless securities with deterministic returns $r_t^0$. The VAR$(p)$ model series for the $\mathbf{R}_t = (r_t^1,\cdots,r_t^n)'$ is given by:
\begin{equation}
    \mathbf{R}_t = \bm{\mu}+\sum_{j=1}^p \bm{\Phi}_i(\mathbf{R}_{t-j} - \bm{\mu})+\bm{\epsilon}_t \hspace{5mm} t=1,\cdots,T,
\end{equation}
for sequence of $(n\times n)$ coefficient matrices $\bm{\Phi}_1,\cdots,\bm{\Phi}_p,$ where $\mathrm{E}[\mathbf{R}_t|\mathcal{F}_t] = \bm{\mu}$ and the $\bm{\epsilon}_t$ is a $(n\times1)$ zero-mean white noise process with invariant covariance matrix $\bm{\mathrm{\Sigma}}$.

We write $\mathbf{R}_t = (r_t^1,\cdots,r_t^n)'$  an $(n\times1)$ vector of time series into multivariate normal dynamic linear model (DLM) as:
\begin{equation}
     \mathbf{R}_t = \bm{\mathrm{\Pi}}'\mathbf{Z}_t + \bm{\epsilon}_t
\end{equation}
where $\mathbf{Z}_t' = (1,\mathbf{R}'_{t-1},\cdots,\mathbf{R}'_{t-p})$, $\bm{\mathrm{\Pi}} = [\pi_1,\cdots,\pi_n]$ and $\pi_i\: (i=1,\cdots,n)$ is a $(k\times 1)$ vector of parameters where $k=np+1$. Consider $\mathbf{Z}$ as a $(T\times k)$ matrix with $t^{th}$ row given by $\mathbf{Z}_t'$. We represent $\bm{\mathrm{\Pi}}$ into a $(nk\times1)$ vector denoted by $vec(\bm{\mathrm{\Pi}})$. For stationary and ergodic VAR models, $vec(\bm{\mathrm{\Pi}})$ is consistent and asymptotically normally distributed with asymptotic covariance matrix
\begin{equation}
    \widehat{avar}(vec(\bm{\mathrm{\Pi}})) = \bm{\mathrm{\Sigma}} \otimes (\mathbf{Z}'\mathbf{Z})^{-1}
\end{equation}
where 
\begin{equation*}
   \bm{\mathrm{\Sigma}} = \frac{1}{T-k}\sum_{t=1}^T \bm{\epsilon}_t\bm{\epsilon}_t' \hspace{5mm}\text{and}\hspace{5mm} \bm{\epsilon}_t =  \mathbf{R}_t -\bm{\mathrm{\Pi}}'\mathbf{Z}_t.
\end{equation*}
Consider the initial information to implement DLM as follows:
$$(\bm{\mathrm{\Pi}}|D_0) \sim \mathcal{N}[\mathbf{m}_0, \mathbf{C}_0].$$
where $\mathbf{m}_0 = \mathrm{E}[\bm{\mathrm{\Pi}}|D_0] = \bm{\mathrm{\Pi}}$ and $\mathbf{C}_0 = \mathrm{Var}[\bm{\mathrm{\Pi}}|D_0]=\widehat{avar}(vec(\bm{\mathrm{\Pi}}))$. \\
Prediction distribution of the vector return series is:
\begin{align}
&(\mathbf{R}_t|D_{t-1})\sim \mathcal{N}[\mathbf{f}_t, \mathbf{Q}_t]\\
\nonumber\text{where, }& \mathbf{f}_t =\mathbf{Z}_t' \mathbf{m}_{t-1},\\
\nonumber & \mathbf{Q}_t = \mathbf{\tilde{Z}}_t'\mathbf{m}_{t-1}\mathbf{\tilde{Z}}_t + \bm{\mathrm{\Sigma}} \hspace{5mm}\text{and}\hspace{5mm}\mathbf{\tilde{Z}}_t = \begin{bmatrix}
\mathbf{Z}_t' & 0\\
0 & \mathbf{Z}_t'
\end{bmatrix}.
\end{align}
Posterior distribution of $\bm{\mathrm{\Pi}}$ at time $t$ is:
\begin{align}
&(\bm{\mathrm{\Pi}}|D_{t})\sim \mathcal{N}[\mathbf{m}_{t}, \mathbf{C}_{t}]\\
\nonumber \text{where, }& \mathbf{m}_{t} = \mathbf{m}_{t-1}+ \mathbf{A}_t (\mathbf{R}_t - \mathbf{f}_t),\\
\nonumber &\mathbf{C}_t = \mathbf{C}_{t-1}-\mathbf{A}_t \mathbf{Q}_t \mathbf{A}_t'\hspace{5mm}\text{and}\hspace{5mm}\mathbf{A}_t= \mathbf{C}_{t-1}\mathbf{Z}_t' \mathbf{Q}_t^{-1}.
\end{align}
Let $\Tilde{\mathbf{R}}_t = (\Tilde{r}_t^1,\cdots,\Tilde{r}_t^n)'$ denote the sequentially updated return series obtained from the prediction distribution. Define the excess return for $i^{th}$ security at period $t+1$ as $e_t^i=\Tilde{r}_t^i-r_t^0$, $\mathbf{e}_t = (e_t^1,\cdots,e_t^n)'$  and the investment strategy $U_t= (u_t^1,\cdots,u_t^n)$ for $i=1,2,\cdots, n$ and $t=0,1,\cdots,T-1$, where $u_t^i$ is the amount invested in the $i^{th}$ risky asset at time $t$. 

The multi-period mean-variance portfolio selection for uncertain exit time with serial and cross correlation among the risky asset can be formulated as:
\begin{equation}
P(\omega)\left\{\begin{array}{ll} \underset{\bm{U}}{\text{max}} \hspace{3mm} \mathrm{E}_{0}(x_{T\wedge\tau})-\omega \mathrm{Var}_0(x_{T\wedge\tau}) \\ \text{s.t.} \hspace{5mm} x_{t+1}=r_t^0 x_t + \mathbf{e}_t U_t, \hspace{5mm}t=0,1,...,T-1,\end{array}\right.
\end{equation}
where $\omega$ is a given positive constant, representing the investor's risk aversion. To obtain the optimal portfolio allocation strategy we have extension of theorem 1 to the situation with multiple assets.

\begin{theorem}\label{thm:4}
The optimal strategy of the mean-variance problem $P(\omega)$ is given by
\begin{equation}\label{eq:(1)}
U_t^* = \frac{1+2\omega\lambda_0 x_0}{2\omega(1-\Theta)}\mathrm{E}_t^{-1}(\omega_{t+1}\mathbf{e}_t \mathbf{e}_t')\mathrm{E}_t(\lambda_{t+1}\mathbf{e}_t) - \mathrm{E}_t^{-1}(\omega_{t+1}\mathbf{e}_t \mathbf{e}_t')\mathrm{E}_t(\omega_{t+1}\mathbf{e}_t) r_t^0 x_t,
\end{equation}
where
\begin{align*}
\xi_t &= \mathrm{E}_t(\lambda_{t+1}\mathbf{e}_t)\mathrm{E}_t^{-1}(\omega_{t+1}\mathbf{e}_t\mathbf{e}_t')\mathrm{E}_t(\lambda_{t+1}\mathbf{e}_t), \hspace{15mm} \Theta = \sum_{t=1}^T \mathrm{E}_0(\xi_{t-1}),\\
\omega_t &= p_t+(r_t^0)^2\left[\mathrm{E}_t(\omega_{t+1}) - \mathrm{E}_t(\omega_{t+1}\mathbf{e}_t)\mathrm{E}_t^{-1}(\omega_{t+1}\mathbf{e}_t\mathbf{e}_t')\mathrm{E}_t(\omega_{t+1}\mathbf{e}_t)\right],\hspace{5mm} \omega_T = p_T,\\
\lambda_t &= p_t+r_t^0\left[\mathrm{E}_t(\lambda_{t+1}) - \mathrm{E}_t(\lambda_{t+1}\mathbf{e}_t)\mathrm{E}_t^{-1}(\omega_{t+1}\mathbf{e}_t\mathbf{e}_t')\mathrm{E}_t(\omega_{t+1}\mathbf{e}_t)\right],\hspace{5mm} \lambda_T = p_T,
\end{align*}
for $t=0,1,...,T-1.$

The efficient frontier of the original problem $P(\omega)$ is given by
\begin{equation}
\mathrm{Var}_0(x_{T\wedge\tau})=\frac{(1-\Theta)}{\Theta}\left[\mathrm{E}_0(x_{T\wedge\tau})-\frac{\lambda_0 x_0}{1-\Theta}\right]^2 + \left[\omega_0-\frac{\lambda_0^2}{1-\Theta}\right]^2x_0^2, 
\end{equation}
for $\hspace{5mm} \mathrm{E}_0(x_{T\wedge\tau}) \in \left[\frac{\lambda_0 x_0}{1-\Theta},+\infty\right).$
\end{theorem}

\section{Simulations}
\label{sec:5}
In this section, we consider a fixed time series model for asset returns to study the effect of model parameters on the efficient frontier. We have developed the theory for VAR($p$) models for multiple assets. Here we restrict to $p=1$ for a single asset to better interpret the results. Consider the AR(1) model for returns:
\begin{equation}
(r_t-\mu)=\phi(r_{t-1} -\mu)+\epsilon_t  \label{eq:(3)}
\end{equation}
Here, $\phi$ captures the autocorrelation in the return series, $\mu$ is the unconditional expectation of $r_t$ and $\epsilon_t$ is a random variable having normal distribution with mean 0, variance $\sigma^2$ and $\epsilon_t$ is independent of $\epsilon_s (s<t)$. We investigated the effect of each of these three AR model parameters on the return series.

Keeping deterministic and random factors constant, we simulated 100 samples of the model mentioned above for $\phi$ ranging from $-0.9$ to $0.9$. Table \ref{tab:1} presents the probability when sequentially updating using the Bayesian technique performs better for the range of values of $\phi$. For $|\phi|$ close to 1, the autoregressive model breaks down due to non-stationarity. In this case, to capture the dependence, we need to take into account the integrating effect. Bayesian updating performs best for $\phi \in [-0.4, 0.4]$ as the serial correlation among asset returns is adequately captured. However, when $\phi =0$, the probability is quite low as there is no autocorrelation among asset returns and both the methods perform equivalently.

\begin{table}
    \caption{Probability that bayesian updates perform better for different values of $\phi$ (Fix $\mu = 0.01$ and $\sigma^2 = 0.2$)}
    \label{tab:1}
    \begin{tabular}{ll|ll} 
        \hline\noalign{\smallskip}
    Value of $\phi$ & Probability & Value of $\phi$ & Probability  \\
      \noalign{\smallskip}\hline\noalign{\smallskip}
      -0.9 & 0.06 & 0.1 & 1\\
      -0.8 & 0.13 & 0.2 & 1\\
      -0.7 & 0.17 & 0.3 & 0.99\\
      -0.6 & 0.36 & 0.4 & 0.82\\
      -0.5 & 0.54 & 0.5 & 0.53\\
      -0.4 & 0.83 & 0.6 & 0.38\\
      -0.3 & 0.98 & 0.7 & 0.25\\
      -0.2 & 0.99 & 0.8 & 0.05\\
      -0.1 & 0.99 & 0.9 & 0.01\\   
      	0  & 0.17 &     &     \\ 
      	    \noalign{\smallskip}\hline  
    \end{tabular}
\end{table}

Another model parameter is $\mu$, which is deterministic in nature. Table \ref{tab:2} shows the probability when dynamic programming with updates perform better for different values of deterministic factor $\mu$ and fixing other parameters. As the absolute value of deterministic factor $|\mu|$ increases, the relative contribution of the dependent part decreases, and both the methods are reasonably comparable. But for small values of $|\mu|$, the effect of dependence factor $\phi$ dominates and leads to better performance of the proposed method at the same level of variance.

\begin{table}
\caption{Probability that bayesian updates perform better for different values of $\mu$ (Fix $\phi = 0.1$ and $\sigma^2 =1$)}
 \label{tab:2}
    \begin{tabular}{ll|ll} 
        \hline\noalign{\smallskip}
    Value of $\mu$ & Probability & Value of $\mu$ & Probability  \\
         \noalign{\smallskip}\hline\noalign{\smallskip}
      -0.5 & 0.07 & 0.01 & 1\\
      -0.4 & 0.32 & 0.03 & 1\\
      -0.3 & 0.43 & 0.05 & 0.95\\
      -0.2 & 0.54 & 0.1 & 0.83\\
      -0.1 & 0.78 & 0.2 & 0.80\\
      -0.05 & 0.99 & 0.3 & 0.51\\
      -0.03 & 0.99 & 0.4 & 0.43\\
      -0.01 & 1 & 0.5 & 0.09\\   
      	0  & 1 &     &     \\   
      	    \noalign{\smallskip}\hline
    \end{tabular}
    
\end{table}

Finally, we fix deterministic and dependence factors and vary the variance of random part i.e., $\epsilon_t$. From Table \ref{tab:3}, it is clear that both the models perform adequately for a lower variance of $\epsilon_t$, as in lesser variance, there is not much gain in updating the parameters. However, for higher values of variance our algorithm always performs better.

\begin{table}
   \caption{Probability that bayesian updates perform better for different values of $\sigma^2$ (Fix $\phi = 0.1$ and $\mu =0.01$)}
    \label{tab:3}
    \begin{tabular}{ll} 
    \hline\noalign{\smallskip}
    Value of $\sigma^2$ & Probability \\
      \noalign{\smallskip}\hline\noalign{\smallskip}
      0.05 & 0.74\\
      0.1 & 0.80\\
      0.15 & 0.96\\   
      0.2  & 0.99\\   
      0.25  & 1\\   
      \noalign{\smallskip}\hline
    \end{tabular}
\end{table}

\section{Empirical Illustration}
\label{sec:6}
\subsection{Data}
\label{subsec:1}
For an empirical illustration, we apply the modeling and optimization framework to the US stock market index S\&P500 and five stocks from it based on market capitalization which include Apple Inc.(AAPL), Microsoft Corporation(MFST), Amazon.com Inc.(AMZN), Facebook Inc. Class A(FB) and Alphabet Inc. Class A(GOOGL), corresponding to one year up to two years. An asset weekly returns $r_t$ is given by $(P_t/P_{t-1})$ where $P_t$ denotes the asset price at time $t$. We downloaded our data from Yahoo finance in USD. The time period of the return data is 156 weeks from July 2017 to June 2020. We used an initial period of 130 weeks of this data for estimating the parameters of the autoregressive process, followed by sequential updating of returns using the remaining data set by applying the dynamic linear model.

\subsection{Results}
\label{subsec:2}
 In this section, we provide a numerical example to demonstrate the impacts of the uncertainty of exit time and the serial correlations of returns on the efficient frontier. To illustrate the results, we have considered a single risky asset S\&P500. We assume that the return rate $r_t$ is subject to AR($p$) model. The order of the AR model is estimated by taking different combinations for the lag values, ranging from one to five, and selecting the most appropriate based on the AIC values. The results show that the return rate is well-captured by AR($1$) model given in equation (\ref{eq:(3)}).

To illustrate the effect of Bayesian updating in multiperiod, we compare the efficient frontiers of mean-variance portfolio optimization problem with and without the dynamically updating the returns. For the above mentioned data, AR parameters estimate for initial time $t = 0$ are obtained as: $\mu_0 = 0.00207$, $\phi_0 = -0.13366$ and $\sigma^2_0 = 0.00085$. Other parameters are set as: initial wealth $x_0=1$, length of investment time $T=26$, i.e. 26 weeks, riskless asset return $r_t^0 = 0.0057$ (risk-free rate for USA) for $t=1,2,...,T$ and exit time distribution is given as:
$$P\{T\wedge\tau =t\} =\left\{ \begin{array}{ll} 0.001 &   t=1,...,T-1\\ 1-0.001(T-1) & t=T  \end{array}\right.$$

\begin{figure}
  \centering
    \includegraphics[width=\textwidth]{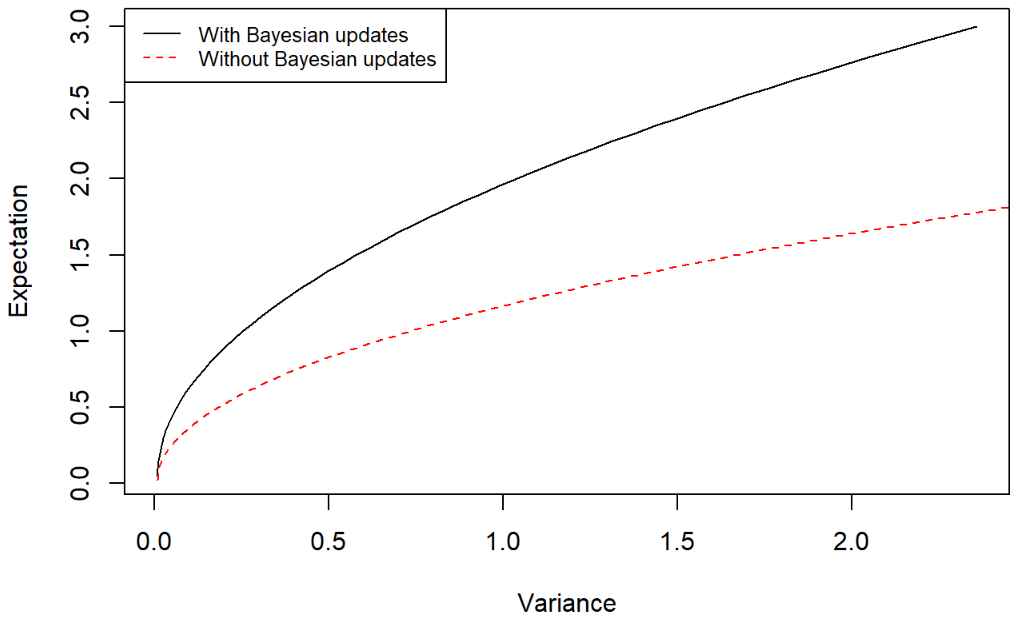}
   \caption{Efficient Frontiers of updated and non-updated model}
   \label{fig:1}
\end{figure}

\begin{figure}
  \centering
    \includegraphics[width=\textwidth]{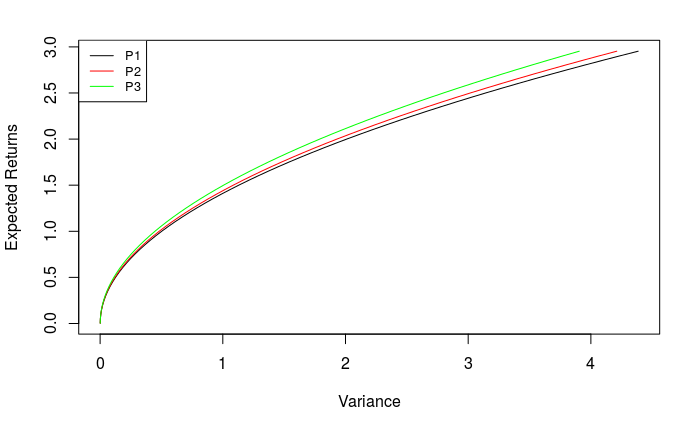}
   \caption{Efficient frontiers with different exit time distribution}
   \label{fig:3}
\end{figure}

\begin{figure}
\centering
    \includegraphics[width=\textwidth]{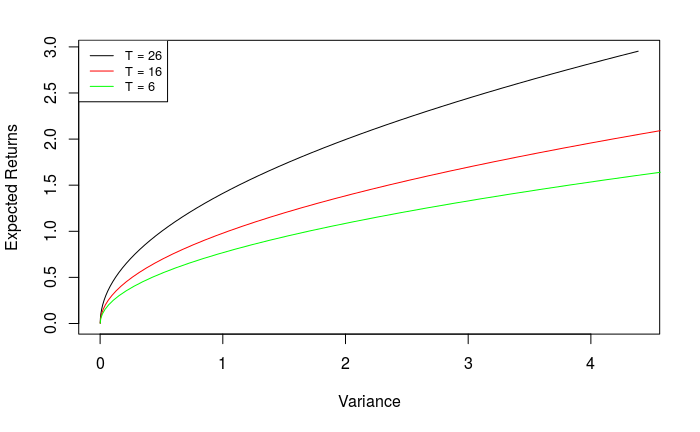}
    \caption{Efficient frontiers with different exit time}
    \label{fig:4}
\end{figure}

Figure \ref{fig:1} illustrates that it is beneficial for the investor to rebalance the portfolio by looking at the current prices and subsequently update the model parameters. We also compared the efficient frontiers of a dynamic mean-variance optimization problem with bayesian updates under different exit time and different exit time distribution.  Probability distributions of uncertain exit time $t=T\wedge\tau$ by varying spread considered for the study are
\begin{equation}
P_1 \{T\wedge\tau =t\} =\left\{ \begin{array}{ll} 0.001 & t=1,...,25\\ 0.975 & t=26  \end{array}\right.
\end{equation}
\begin{equation}
P_2 \{T\wedge\tau =t\} =\left\{ \begin{array}{ll} 0.001 & t=1,...,24\\ 0.3 & t=25 \\ 0.676 & t=26  \end{array}\right.
\end{equation}
\begin{equation}
P_3 \{T\wedge\tau =t\} =\left\{ \begin{array}{ll} 0.001 & t=1,...,15\\ 0.05 & t=16,...,25 \\ 0.485 & t=26  \end{array}\right.
\end{equation}
The broader spread of exit time distribution offers investors more number of opportunities to exit the market and earn the maximum gain when the market is in his favor, as illustrated in Figure \ref{fig:3}.
Figure \ref{fig:4} compares the efficient frontier with different exit time. Figure \ref{fig:4} implies that the investor who spends more time in the market, enjoys more expected wealth at the same level of risk than the one with a higher probability of leaving the market at an earlier stage.

\subsection{Credible Interval Prediction}
\label{subsec:3}
\begin{figure}
  \centering
  \begin{minipage}[b]{0.45\textwidth}
    \includegraphics[width=\textwidth]{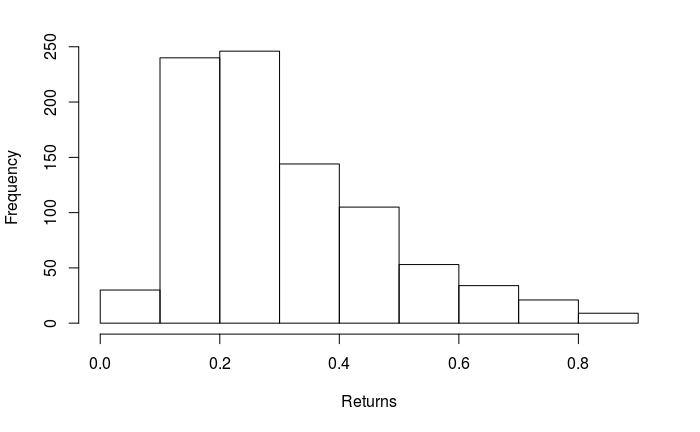}
   \caption{Histogram of empirical distribution of optimal portfolio returns for $\omega=3$}
   \label{fig:5}
  \end{minipage}
  \hfill
  \begin{minipage}[b]{0.45\textwidth}
    \includegraphics[width=\textwidth]{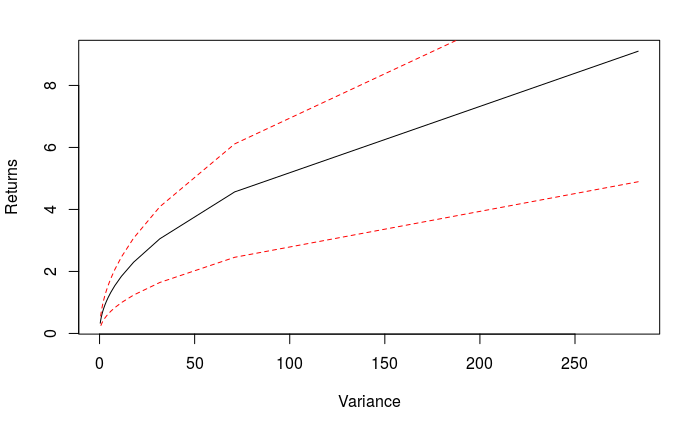}
    \caption{Central 50\% credible intervals for the return of optimal portfolios}
    \label{fig:6}
  \end{minipage}
\end{figure}

In the case of the Bayesian approach, we have the whole prediction distribution of the asset returns instead of just point estimates. Using this prediction distribution, we can form prediction intervals for the expected terminal wealth. To obtain the prediction intervals, we simulate the value of optimal portfolio return for several values of risk tolerance factor $\omega\in(0,3)$  (see Figure \ref{fig:6}).

The prediction intervals for the whole range of $\omega$ are obtained as follows:
\begin{itemize}
\item[(i)] Firstly, obtain the initial information for the AR parameters from the given data set. Using these initial parameters and the recent data, generate the AR($p$) model for asset returns $r_t^i$ as described in (\ref{eq:(5)})  for $t=1,2,...,T$ and $i=1,...,1000$.
\item[(ii)] Fix $\omega$ and for $k \in \{1,...,1000\}$ calculate:
    \begin{itemize}
    \item[(a)] the mean and mean square for excess returns for $t=1,2,...,T$ as follows
    $$\mathrm{E}[e_t^{(k)}]=m_0r_t^{(k)}-r_t^0$$
    $$\mathrm{E}[(e_t^{(k)})^2]=(r_t^0)^2+m_0^2(r_t^{(k)})^2 - 2m_0r_t^{(k)}r_t^0 + \sigma_t^2$$
    \item[(b)] the parameters $p_t,\omega_t,\lambda_t,\theta_t$ and $\Theta$ for $t=1,2,...,T$ using theorem 1.
    \item[(c)] $\lambda^*$ as given (\ref{eq13}) and substitute it in (\ref{eq12}) to obtain optimal portfolio return $r^{*(k)}$.
    \end{itemize}
\item[(iii)] Obtain the empirical distribution of the optimal portfolio return using these simulated values $r^{*(k)},\: k = 1,\cdots,1000$, which is represented by the histogram (see Figure \ref{fig:5}). 
\item[(iv)] Fix the significance level of the prediction interval as $\alpha$ and compute the expectation $E_{\omega}[r^*]$, $\alpha/2-$ and $(1-\alpha/2)-$ quantiles from the empirical distribution.
\item[(v)] Use the expectation of optimal portfolio returns to calculate the variance of terminal wealth $Var_{\omega}[r^*]$ using (\ref{eq4}).
\item[(vi)] Repeat steps (ii)-(v) for $\omega \in (0,3]$ to obtain the expectation, variance and intervals for the whole range of $\omega$. Plot the expectation along with prediction intervals from (iv) for the computed values of variance in part (v).
\end{itemize}
A portfolio with a higher risk aversion coefficient is less risky and therefore lies more left on the efficient frontier. Thus, the smaller $\omega$ will have large credible intervals. 

\subsection{Results for Multiple Risky Assets}
\begin{figure}
  \centering
    \includegraphics[width=\textwidth]{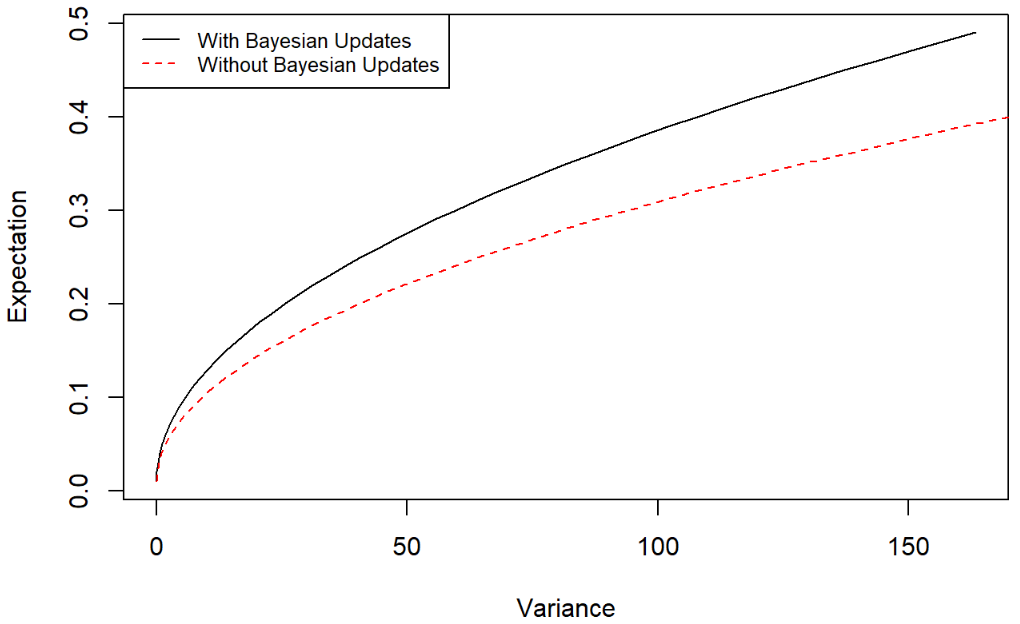}
   \caption{Efficient frontiers for multiple assets of Bayesian updated and non-updated model}
   \label{fig:7}
\end{figure}

To demonstrate the impacts of the serial and cross-correlations among returns of various assets on the efficient frontier, we consider five stocks mentioned in Sect. 6.1 for our portfolio. We assume that the return rate matrix $\mathbf{R}_t$ is subject to VAR(p) model. Based on the AIC values VAR(1) model well-capture the interdependencies among the chosen stocks. Thus, the model we obtained is as follows:
\begin{equation}
    \mathbf{R}_t = \bm{\mu} + \bm{\Phi}(\mathbf{R}_{t-j} - \bm{\mu})+\bm{\epsilon}_t \hspace{5mm} t=1,\cdots,T,
\end{equation}
where $\mathrm{E}[\mathbf{R}_t|\mathcal{F}_t] = \bm{\mu}$, $\bm{\Phi}$ is a $(5\times 5)$ coefficient matrices and the $\bm{\epsilon}_t$ is a $(5\times1)$ zero-mean white noise process with invariant covariance matrix $\bm{\Sigma}$. VAR parameters estimate for initial time $t= 0$ are obtained as:
\begin{equation*}
    \bm{\mu}_0 = \begin{bmatrix}
 0.006632  \\ -0.01047 \\ 0.00704 \\ 0.00245 \\ 0.00299
\end{bmatrix}\: \text{and}\:
\bm{\Phi}_0 = \begin{bmatrix}
 0.0399 & 0.2395 & -0.0185 & 0.0322 & 0.1183\\
0.0160 & -0.1943 & -0.0006 & -0.0020 &  0.0093\\
0.0684 & -1.1354 & 0.0137 & 0.1444 & -0.0398\\
 -0.0618 & 2.7002 & 0.0131 & -0.1892 & -0.0411\\
-0.0882 & -2.1994 & 0.0667 &  0.0850 & -0.0589
\end{bmatrix}.
\end{equation*}
Other parameters are similar to the single risky asset case. Figure \ref{fig:7} illustrates that our model with Bayesian forecasting captures the cross-correlation among assets and reduces the risk for the investors. We also observe that the variance is quite notable in the case of multiple assets model. The reason for this significant variance is low diversification provided by only five stocks compared to S\&P500. Our portfolio consists of "Big Five" American technology companies belonging to the same industry leads to a further reduction in diversification.

\section{Conclusion}
\label{sec:7}
To solve the mean-variance analysis in multidimensional, we considered both the drift vector and covariance matrix of asset returns as uncertain and assumed to have prior distribution based on the past data. This prior information and Bayesian forecasting method are used to obtain the posterior distribution of asset returns at each time point. The dynamic programming approach and the embedding technique of Li and Ng \cite{li2000optimal} are applied to these predicted returns. This approach of combining the Bayesian technique with dynamic programming allows us to find optimal portfolio weights based on historical as well as the latest available information of asset returns. In particular, we showed that the constructed Bayesian efficient frontier results in higher expected returns at a similar level of risk, presented by a numerical example. Another advantage of the Bayesian approach is that it allows us to construct the prediction interval of future realizations of the optimal portfolio returns. The numerical example also shows that the exit time have significant impacts on the optimal strategy and the efficient frontier. Simulation of the AR(1) process allows us to give more importance to the recent data and showed that the dynamic programming with the Bayesian approach efficiently captures the serial correlation of asset returns.

\section*{Appendix}
\label{sec:8}
Instead of solving the orginial problem $P(\omega)$  directly, we first consider the transformed auxiliary problem:
$$A(\lambda,\omega)\left\{ \begin{array}{ll} \underset{U}{\text{max}} \hspace{3mm} E_0(\lambda x_{T\wedge\tau}-\omega x_{T\wedge\tau}^2) \\ \text{s.t.} \hspace{5mm} x_{t+1}=r_t^0 x_t + e_t U_t, \hspace{5mm}t=0,1,...,T-1,\end{array}\right.$$
for a given constant $\lambda>0$. 
The following two theorems can be proven by similar method described by Li and Ng \cite{li2000optimal}.
\begin{theorem}\label{thm:1}
Any optimal solution of $P(\omega)$ will be optimal solution of $A(\lambda,\omega)$  with $\lambda^* = 1+ 2\omega E_0(x_{T\wedge\tau})|_{U^*}$.
\end{theorem}

\begin{theorem} \label{thm:2}
If $U^*$ is the optimal solution of $A(\lambda,\omega)$ then the necessary condition for $U^*$ to be the optimal solution of $P(\omega)$ is $\lambda^* = 1+ 2\omega E_0(x_{T\wedge\tau})|_{U^*}$.
\end{theorem}

Zhang and Li \cite{zhang2012multi} derived the analytical solution of the auxiliary problem by transforming the auxiliary problem $A(\lambda,\omega)$ into portfolio selection problem with certain exit time and applied dynamic programming approach to solve it, which is presented in the following theorem.

\begin{theorem}\label{thm:3}
The value function of problem $A(\lambda,\omega)$ is
\begin{align*}
f_t*(x_t) &=  \max_{u_t} f_t(x_t)\\
&=\max_{u_t}E_t\left[\sum_{s=t}^T (\lambda x_s-\omega x_s^2)p_s\right]\\
&=-\omega \omega_tx_t^2+\lambda \lambda_t x_t + \Xi, \hspace{10mm} t=0,1,...,T-1,
\end{align*}
and the optimal strategy of problem $A(\lambda,\omega)$ is given by
$$u_t*=\frac{\lambda}{2\omega}\frac{E_t(\lambda_{t+1}e_t)}{E_t(\omega_{t+1}e_t^2)}-\frac{E_t(\omega_{t+1}e_t)}{E_t(\omega_{t+1}e_t^2)}r_t^0 x_t, \hspace{10mm} t=0,1,...,T-1,$$
where \begin{align*}
\Xi_t & = E_t\left[\frac{\lambda^2}{4\omega}\sum_{s=t}^{T-1} \theta_s\right],\\
\omega_t& = p_t+(r_t^0)^2\left[E_t(\omega_{t+1}) - \frac{E_t^2(\omega_{t+1}e_t)}{E_t(\omega_{t+1}e_t^2)}\right], \hspace{5mm} \omega_T = p_T,\\
\lambda_t & = p_t+r_t^0\left[E_t(\lambda_{t+1}) - \frac{E_t(\omega_{t+1}e_t)E_t(\lambda_{t+1}e_t)}{E_t(\omega_{t+1}e_t^2)}\right], \hspace{5mm} \lambda_T = p_T,
\end{align*}
for $t=0,1,...,T-1.$
\end{theorem}

Now, insert the optimal strategy given in Theorem \ref{thm:3} into wealth dynamics described in equation (\ref{eq1}) for $t=T$ and taking expectations on both sides based on the information available at time $T-1$, we obtain
\begin{equation}\label{eq11}
E_{T-1}(\lambda_T x_T) =\lambda_{T-1}x_{T-1}-p_{T-1}x_{T-1}+\frac{\lambda}{2\omega}\theta_{T-1}.
\end{equation}
Taking expectations recursively on both sides of equation \ref{eq11} at time $T-2,...,1,0$, we conclude that
\begin{equation}\label{eq12}
E_0(x_{T\wedge\tau})=\lambda_0x_0+\frac{\lambda}{2\omega}\Theta.
\end{equation}
Substituting (\ref{eq12}) in Theorem \ref{thm:2} to satisfy the necessary condition of attaining the optimality of the original problem $P(\omega)$ for the same optimal strategy of problem $A(\lambda,\omega)$. We obtain
\begin{equation}\label{eq13}
\lambda^* = \frac{1+2\omega\lambda_0x_0}{1-\Theta}.
\end{equation}
Finally, subsitute $\lambda^*$ in Theorem \ref{thm:3} to yield the optimal strategy for $P(\omega)$ and efficient frontier, which is summarized in the Theorem \ref{thm:4}.

%
%


\bibliographystyle{spmpsci}      
\bibliography{paper}   

\end{document}